\def\bfrm{{\bf}}
\def\th{{\thinspace}}
\def\ngth{{\negthinspace}}

\def\lgt{ \raise4pt \hbox{$<$}\kern-9pt\lower1.5pt \hbox{$>$}}
\def\glt{\raise4pt \hbox{$<$}\kern-9pt\lower1.5pt\hbox{$>$}}
\def\approxgt{\th{\raise3pt\hbox{${\scriptstyle>}$}
           \kern-6pt\lower1.1pt\hbox{${\scriptstyle\sim}$}}\th}
\def\approxlt{\th{\raise3pt\hbox{${\scriptstyle<}$}
           \kern-6pt\lower1.1pt\hbox{${\scriptstyle\sim}$}}\th}
\def\ni{\noindent}

\documentstyle[12pt,world_sci]{article}
\input psfig

\begin{document}

\title{STELLAR ARRHYTHMIAS}

\author{J. ROBERT BUCHLER\\
{\em Physics Department, University of Florida}\\
{\em Gainesville FL  32611, USA}}

\maketitle

\begin{abstract}

 The light-curves of the two large amplitude variable stars, R Scuti and AC
Herculis, display irregular pulsation cycles with 'periods' of 75 and 35 days,
respectively.  We review the evidence that the observed time-series are
generated by low dimensional chaotic dynamics.  In particular, a global flow
reconstruction technique indicates that the minimum embedding dimensions are 4
and 3, for R~Sct and AC~Her, respectively, whereas the fractal dimensions are
inferred to be $d_L\approx 3.1$ and $2.05 \approxlt d_L \approxlt 2.45$,
respectively.  It thus appears that the dimensions of the dynamics themselves
are also 4 and 3.

\end{abstract}

\section{Introduction}

For several centuries it has been known that a sizeable subset of the stars
exhibits a variable energy output (luminosity).  The best known and best
studied of these variable stars are the Cepheids that have a clock-like
periodicity.  Their notoriety comes from the tight relation between their
period and luminosity and their consequent role as cosmological distance
indicators.  In parallel, there exists a class of variable stars with
light-curves that instead exhibit various degrees of irregularity.  These
objects undergo changes in luminosity that can be as large as a factor of
forty, indicating violent radial (i.e. $\ell = 0$) pulsations.  Very little
theoretical attention had been devoted to these stars.  

Why do we think that the dynamics underlying these irregular pulsations could
be low dimensional and chaotic?  One of the reasons is that decade old
numerical hydrodynamical simulations showed low dimensional chaotic behavior
and period doubling cascades in models.\cite{PDletter,PDpaper} However, the
observational confirmation of low dimensional chaos in these stars, that is
the subject of this paper, had to wait a little longer for a number of reasons.

\section{The observational data}

Nonlinear techniques are new in variable star Astronomy, and it is therefore
hard to find data sets that are suitable for a modern nonlinear time series
analysis and the search for low dimensional chaos.  Partially because of the
long 'periods' involved, typically tens to hundreds of days, rarely are enough
pulsation cycles covered, or they are covered incompletely.  Generally,
professional astronomers have not seen any merit in dedicating the necessary
telescope time to what the consider the gathering of 'useless' erratic data.
In contrast, thanks to the observations of numerous amateur astronomers, the
American Association of Variable Star Observers (AAVSO) has recently made
available long and relatively well sampled light-curve data of two stars, R
Scuti and AC Herculis.  (In Astronomy variable stars were named after the
constellation in which they appear preceded by one or two capital letters,
until the letters were exhausted).  Both objects are characterized by irregular
cycles of $\approx$ 75\th d and 35\th d, respectively, with alternating shallow
and deep minima.  The details of the analyses of these light-curves have been
published elsewhere.\cite{PRlett,s3,s4} Here we present a summary review.

 \centerline{\vspace{10pt}\psfig{figure=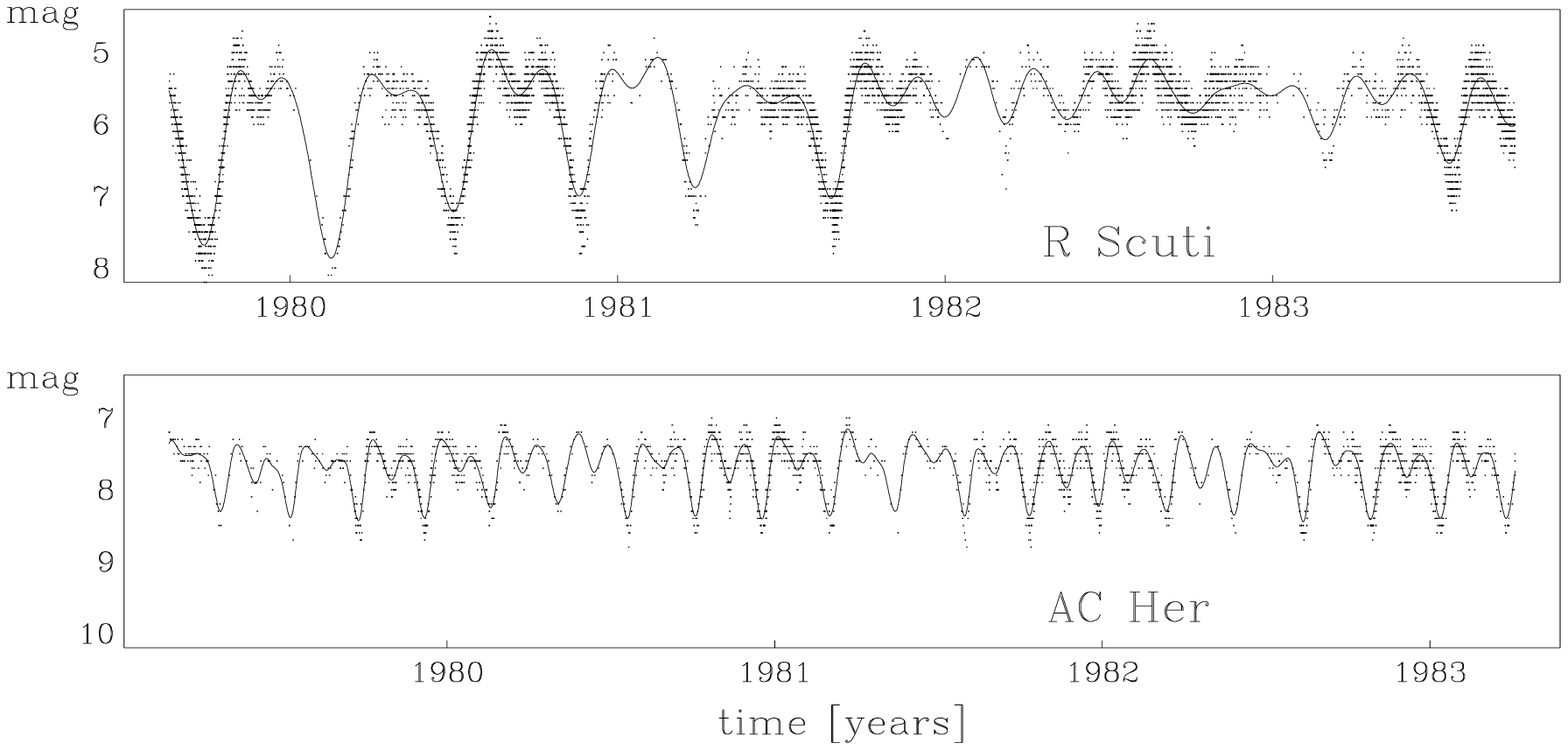,width=15cm}}
 \vspace{-5mm}
 \ni{\footnotesize Fig.~1\ 
 Typical segments of observational lightcurve data (dots) and fits (lines),
magnitude {\sl versus} time [years]; R~Scuti (top) and AC~Herculis (bottom).
 \vspace{6mm}
 }

Fig.~1 displays typical subsections of the two data sets (points) with the
solid line showing our smoothed and interpolated fit (see below).  These data
have been taken visually by many amateur observers worldwide at essentially
random times.  Therefore there is considerable scatter in the data and there
are some observational gaps.  We have plotted the lightcurves in magnitudes
(for historical reasons defined as $-2.5~{\rm Log}\times$ luminosity) rather
than the more 'physical' luminosity.  For both data sets the error in the
magnitudes is found to have a normal distribution, independent of the magnitude
(which is not true for the luminosity itself).  The reason is to be found in
the visual nature of the observations and the logarithmic response of the human
eye.  We thus work with the magnitude.  Although the individual observations
have a large error associated with them (0.20 and 0.15 mag, in R~Sct and
AC~Her, resp.)  it is possible to extract a reasonably accurate light-curve by
averaging and interpolating.  The individual errors are larger for R~Sct, but
there are about four times as many data points and the pulsation amplitude is
about a factor of 4 larger, so that the error on the final interpolated curve
is much smaller for R~Sct than for AC~Her.

 \centerline{\psfig{figure=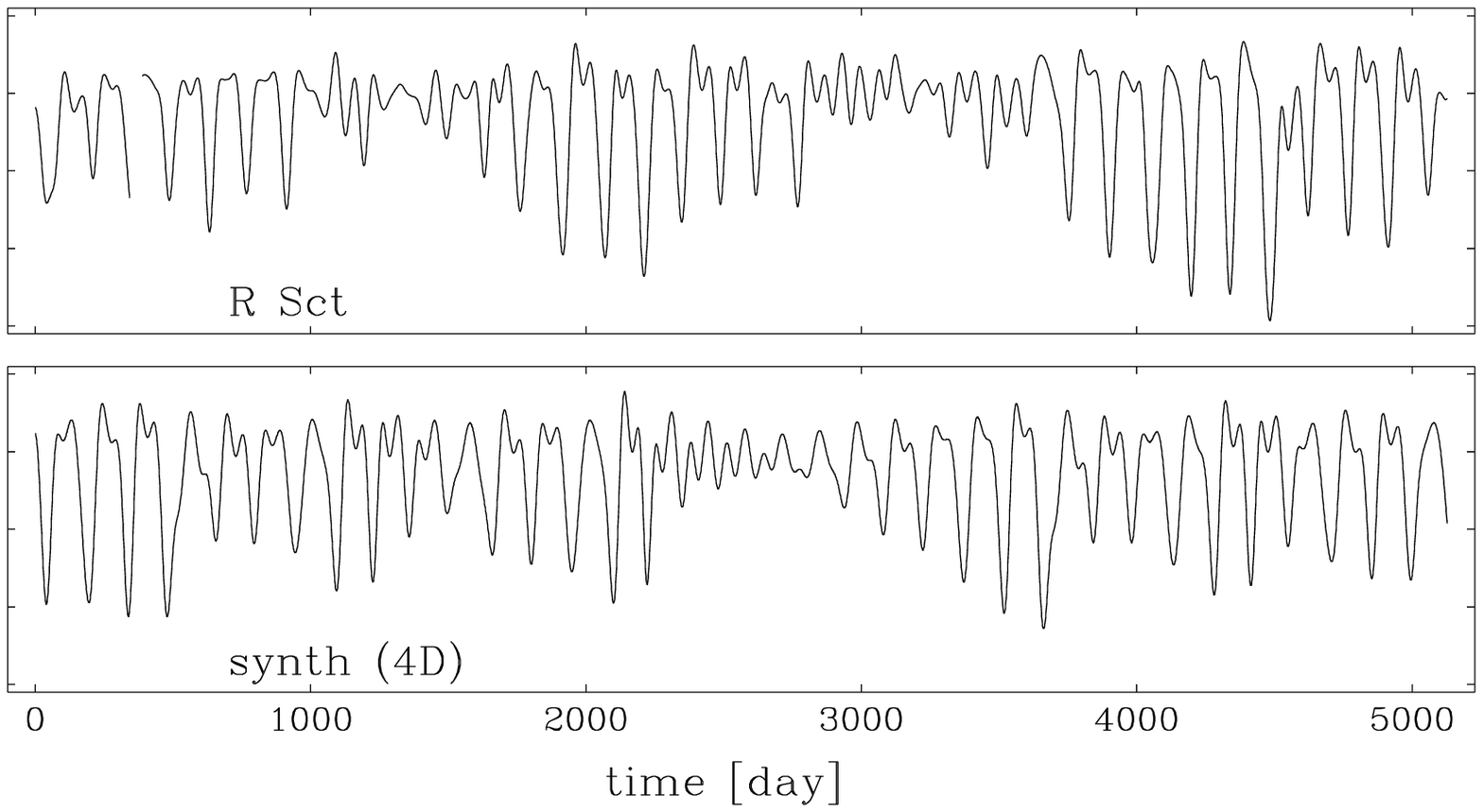,width=14.5cm}}
  \vspace{-3mm}
  \ni{\footnotesize Fig.~2\ 
 Observational lightcurves of R~Scuti and best synthetic signal (4D).
  \vspace{6mm}
 }

 \centerline{\psfig{figure=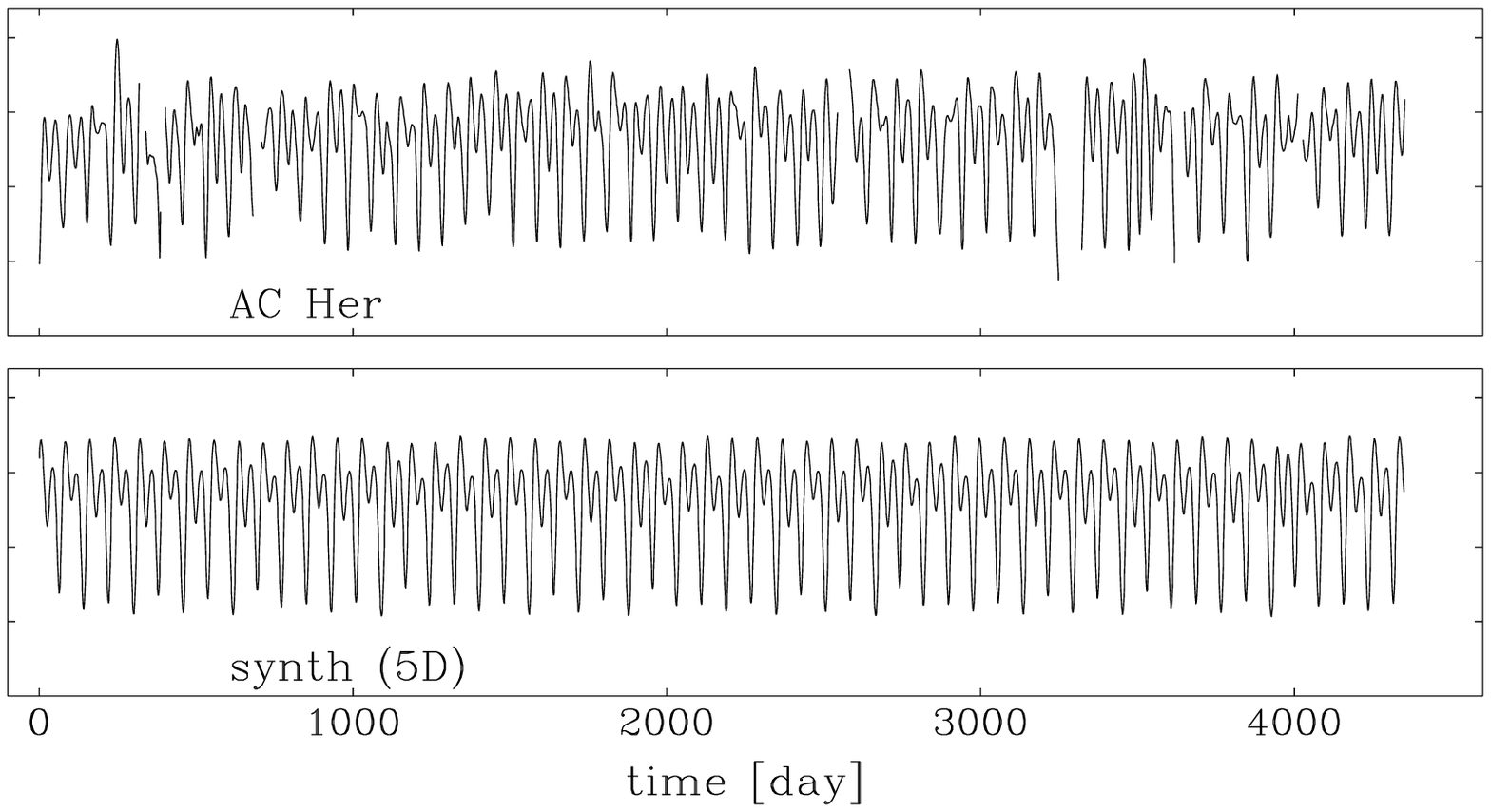,width=14cm}}
  \vspace{-2mm}
 \ni{\footnotesize Fig.~3\ 
 Observational lightcurves of AC~Herculis (top) and best synthetic signal (5D).
 \vspace{6mm}
 }

The preparation of the data is a delicate process.  We want to eliminate as
much observational and other extrinsic noise as possible.  However there is
additional 'intrinsic noise' associated with turbulence and convection in the
star that, strictly speaking, is part of the dynamics.  It is well known that
well developed turbulent motions constitute high-dimensional behavior, and {\sl
a priori} it may be objected this would preclude us from uncovering any low
dimensional dynamics.  The success of our nonlinear analysis provides an {\sl a
posteriori} indication that the pulsation can be separated into a large
amplitude 'true' pulsation, and a low amplitude pulsational jitter.  Our
pre-smoothing eliminates this high-dimensional jitter together with the true
noise, leaving a signal that represents the 'true' low dimensional dynamics.

To be specific, the bare data were first averaged in 5.0 day bins for R~Sct and
in 2.5 day bins for AC~Her, followed by a cubic spline smoothing, where the
smoothing parameter $\sigma$ was a free parameter in the flow reconstruction
below.  The result is then a lightcurve data set that is sampled at equal
time-intervals.

Figs.~2 and 3 in the top rows show the smoothed and interpolated observational
data of R Sct and AC Her, respectively.  In Fig.~4 we display the (normalized)
amplitude Fourier spectra of the two observed lightcurves.  The FS of R~Sct has
much broader features than the more regular AC~Her as is to be expected from
the appearance of the lightcurves.

 \centerline{\psfig{figure=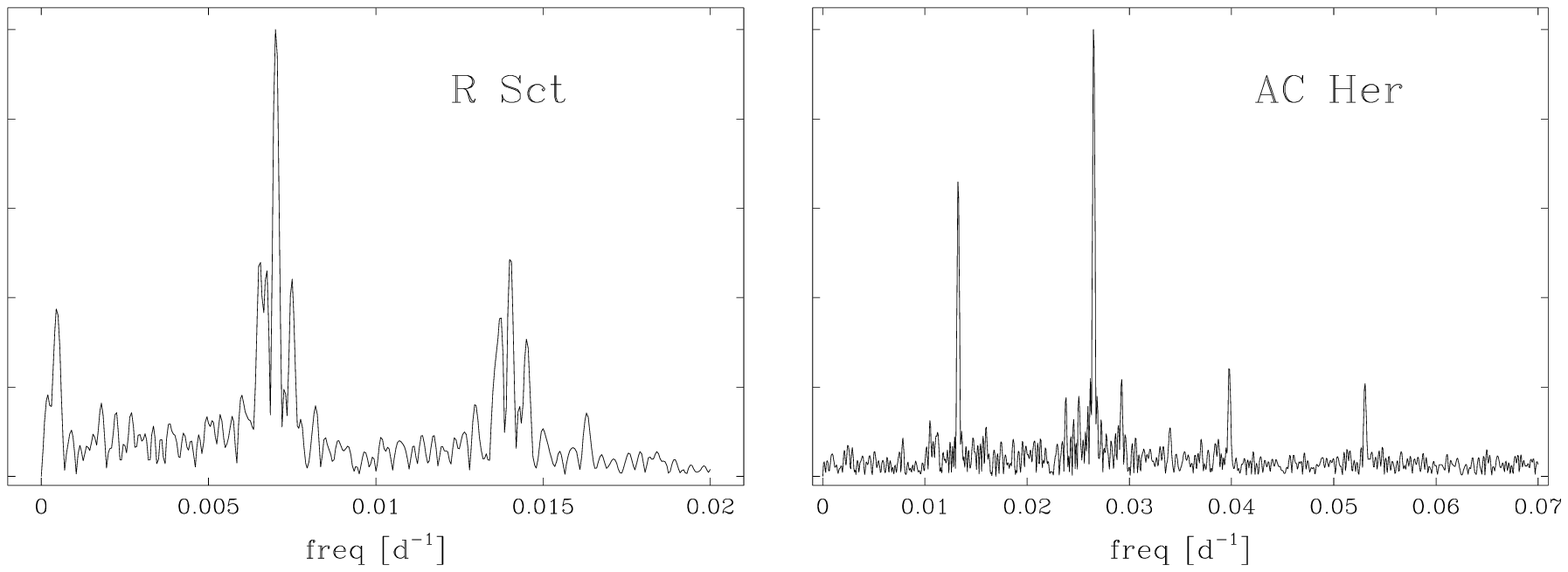,width=17cm}}
 \vspace{-3mm}
\ni{\footnotesize Fig.~4\ 
Amplitude Fourier spectrum of R~Sct and AC~Her lightcurve data.
  \vspace{7mm}
 }

It is also of interest to look at the Broomhead--King (BK)
projections\cite{Abar} of a signal.  These projections onto the eigenvectors of
the correlation matrix are optimal for spreading out the features of the
signal.  The lowest BK projections for R~Sct and AC~Her are thus shown in
columns 1 and 5 in Fig.~5.

 \vspace{2mm}

Before proceeding with a nonlinear analysis aimed at discovering and describing
low dimensional chaos we briefly mention further tests to ascertain that
the signals cannot be the result of a multi-periodic (n-torus) dynamics.

For R~Sct we have performed the following tests.\cite{s3} Taking the
frequencies of the highest 35 peaks in the FS (Fig.~2) we have made a
multi-periodic Fourier fit for the 35 amplitudes and phases.  The fit is quite
good over the temporal range of the data set, but when it is {\sl extrapolated}
in time the signal totally fails to resemble the R~Sct lightcurve.  Similarly,
a linear autoregressive process with white noise (AR or ARMA) bears little
resemblance to the data.  Both tests indicate that the process that produces
the lightcurve is strongly nonlinear.\cite{Abar,WG}

For AC~Her, the situation is not so clear-cut because of the high noise level.
We note here prewhitening with the frequencies of the 24 highest FS peaks
leaves a noise level that is at least 3--4 times the noise that is expected
from the observational error.\cite{s4}

 \centerline{\psfig{figure=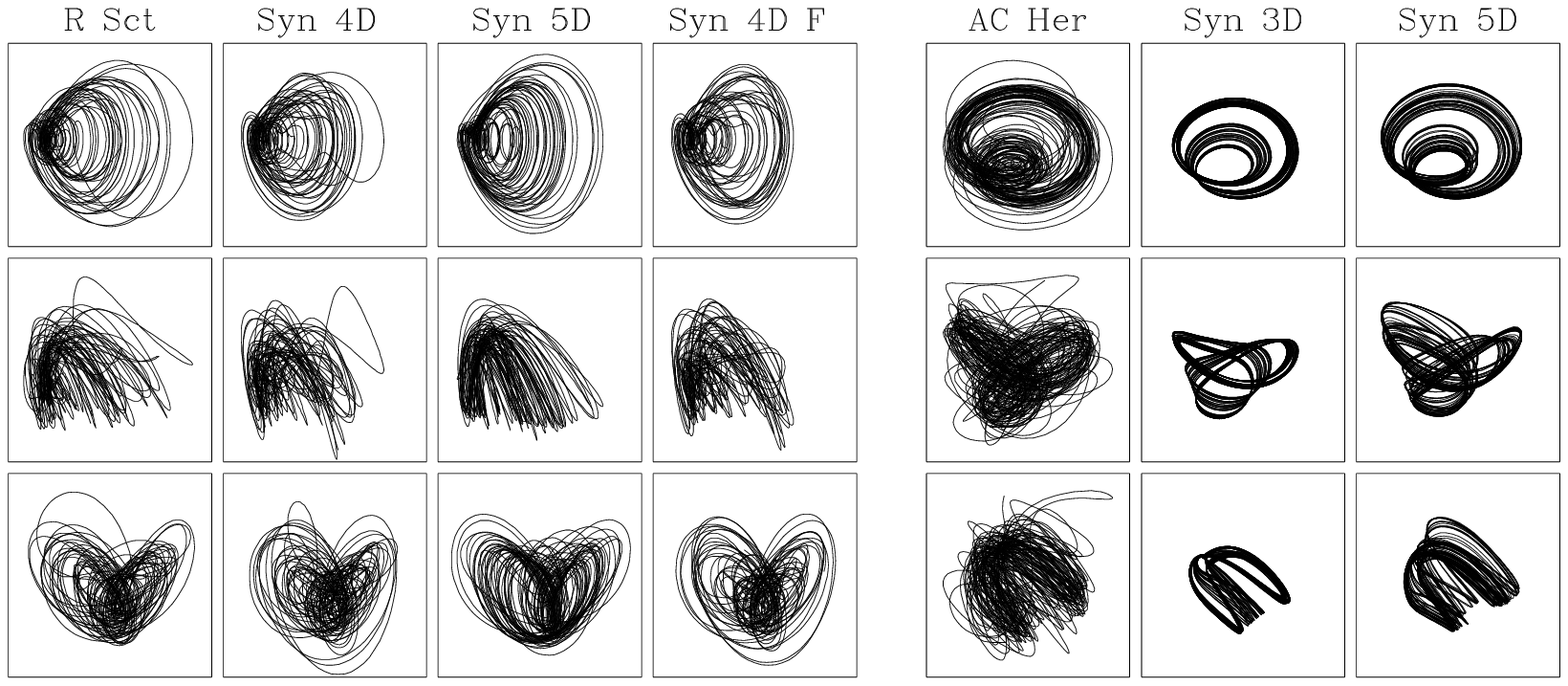,width=15.4cm}}
 \vspace{-4mm} 
 \ni{\footnotesize Fig.~5\
 Broomhead-King projections for the R~Sct and AC~Her data, as well as for the
respective best synthetic signals; from top to bottom, coordinates
($\xi_2$,$\xi_1$),($\xi_3$,$\xi_1$)and ($\xi_3$,$\xi_2$).
 \vspace{6mm} 
 }

There are theoretical reasons for not thinking that so many frequencies could
be observed.  First these stars are believed to be radial pulsators, and the
radial modal spectrum does {\sl not } have a structure similar to that of the
highest peaks of the FS.  Further, even if one insisted on nonradial modes, the
visual observations are necessarily whole disk observations which cannot
possibly resolve nonradial modes with $\ell >2$.  Again the required modal
spectrum would not be compatible with stellar structure.\cite{s4}

The stars cannot be evolving (time-dependent) multiperiodic systems either.
Otherwise there should be a correlation between the multitude of smaller peaks
in the Fourier spectrum when subsequent sections of the data are Fourier
analyzed, which is not the case.
 
\vspace{3mm}

We conclude that neither the R~Sct nor the AC~her lightcurve can reasonably be
explained as being periodic or multi-periodic.

 \vskip 10pt

\section{The global flow reconstruction}

We now make the assumption that the observed lightcurve $s_n=\{s(t_n)\}$ has
been generated by a low dimensional dynamics in some 'physical' phase-space of
{\sl a priori} unknown properties, i.e.
 \begin{equation}
 {\bf dY/dt = G(Y)}
 \end{equation}
 where ${\bf Y}(t)\in {{\bfrm R}}^d$ is the $d$ dimensional state vector of
the system.  We could equivalently assume a stroboscopic description of the
dynamics
 \begin{equation}
 {\bf Y_{n+1} = G(Y_n)}
 \end{equation}
 where ${\bf Y}_n= {\bf Y}(t_n)$.  It is natural to assume that $s(t_n)$ is a
smooth function of the phase-space variable ${\bf Y}$.

A standard reconstruction method starts with the construction of the standard
$d_e$--dim delay vectors\cite{WG,Abar}
 \begin{equation}
 {\bf X_n} = \{s_{n-1},s_{n-2}, \ldots, s_{n_{d_e}}\}
 \end{equation}  
 The {\sl global} flow reconstruction method then assumes that there is a
single function (map) ${\bf F}$ that describes the dynamics all over,
 \begin{equation}
 {\bf X_{n+1} = F(X_n)}
 \end{equation}

 Embedding theorems\cite{Sauer} guarantee that for sufficiently large $d_e$
there is a diffeomorphism between the physical phase-space and the
reconstruction space (the latter is then called an embedding space, {\sl
stricto sensu}).  Following Giona {\sl et al.}\cite{Giona} and
Brown\cite{Brown} we assume that the function ${\bf F(X)}$ is of a multivariate
polynomial form,
 \begin{equation}
 {\bf F(X)}= \sum_i \alpha_i {\cal P}_i({\bf X})
 \end{equation}
 where the sum runs over all polynomials up to degree $p$.  More specifically
we use natural polynomials, i.e. polynomials that are constructed to be
orthogonal on the data set.  However, instead of constructing these polynomials
explicitly by recursion and then determining the coefficients $\alpha_i$
\cite{Brown,PRlett} it is equivalent and much faster to expand the function in
terms of all monomials up to the same order $p$
 \begin{equation}
  {\bf F(X)}= \sum_i \beta_i {\cal M}_i({\bf X})
 \end{equation}
 and then find the expansion coefficients $\beta_i$ through a least squares
minimization of the variance 
 \begin{equation}
 S=\sum_n ||{\bf X_n}-\sum_i \beta_i {\cal M}_i({\bf X_{n-1}})||^2.
 \end{equation}
 For this purpose a SVD (singular value decomposition) approach has been found
to be most stable and efficient.\cite{s2a,s3}

\vspace{2mm}

The global polynomial reconstruction method was first tested\cite{s2a} on the
R\"ossler attractor.\cite{roessler} Using as input a short time-series of only
the first one of the 3 R\"ossler variables one obtains a very good estimation
of the Lyapunov exponents and the fractal dimension of the attractor.
Furthermore it is possible to determine that this scalar signal was generated
by a 3-D flow (R\"ossler's three first order ODEs), in other words it is
possible to determine not only the lowest embedding dimension, but also the
dimension of the actual physical phase-space of the dynamics.

Encouraged by these benchmark results we have applied the reconstruction
method to the observational data of stellar lightcurves shown in Figs.~2 and 3.

Because the observational data sets are so small we have used all the points in
constructing the map in Eq.~7.  We note in passing that in Astronomy we are not
interested in prediction, but rather in uncovering the properties of the
dynamics that underlies the pulsation.  In Fig.~6 we display the behavior of
the error as a function of $d_e$, which therefore is an in-sample error.  The
vertically stacked points are for values of $p$ ranging from 3 to 6.  This
suggests strongly that for R~Sct the minimum embedding dimension is $d_e=4$.
For AC~Her, because of the much larger observational noise, the lowest
achievable error level is seen to be much higher.  Nevertheless Fig.~6 can be
seen to indicate that we may get away with a minimum $d_e$ of 3.  The behavior
of the error norm with dimension by itself is of course insufficient to
establish the minimum embedding dimension.

 \centerline{\psfig{figure=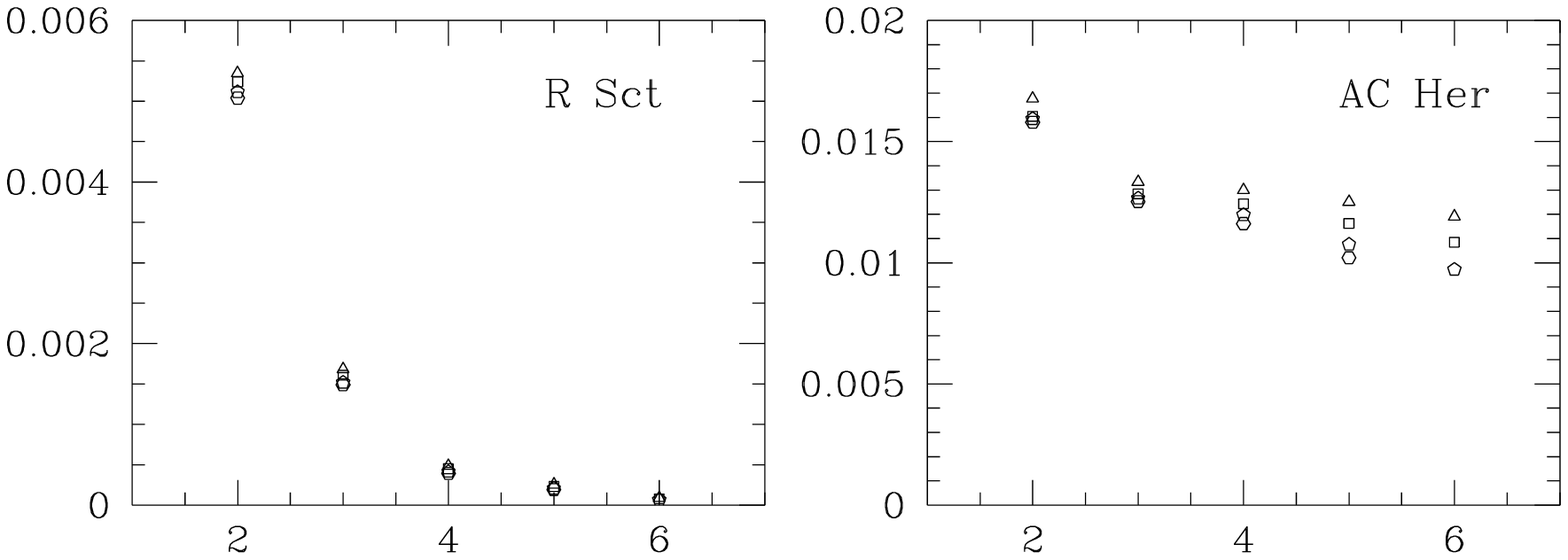,width=16cm}}
 \vspace{-12mm}
 \ni{\footnotesize Fig.~6\
 Behavior of the in-sample error for R~Sct and AC~Her.
 \vspace{6mm}
 }

It is worthwhile noting that the results obtained with the 'false nearest
neighbor' method\cite{Abar} are consistent with these respective minimum
dimensions of 4 and 3.

How do we decide that a map is 'good'?  For that we have to compare the
'synthetic' signals that a map yields to the original data.  (By synthetic
signal we mean data that have been generated by iteration from some seed values
with a map generated according to Eq.~7).  Of course, since we have chaotic
behavior this comparison has to be done in an average or statistical sense.
Unfortunately the observational data set is too small to make statistical
comparison tests very meaningful.  Instead we rely on three rather {\sl ad hoc}
criteria, {\sl viz.} (a) the signals must have the same appearance, (b) their
Fourier spectra must have the same envelope structure and (c) their lowest BK
projections should be similar.  We have found these simple criteria to be quite
useful.  For example in our flow reconstructions it is quite clear that any 3D
maps for R~Sct can be rejected for not being able to capture the dynamics of
the star; the synthetic lightcurves, while they may be chaotic, they bear
essentially no resemblance to the observational data.

In Figs.~2 and 3, bottom rows, we exhibit the synthetic signals we could
produce from what we consider the best of these maps for R~Sct and AC~Her,
respectively.  In the BK projections of Fig.~6 we display the same two signals,
Cols.~2 and 7, respectively.  For comparison we also show synthetic signals for
R~Sct that were obtained with a 5D map (Col.~3).  Col.~6 shows a 3D
reconstruction of AC~Her.  For a comparison of the Fourier spectra we refer to
the original papers.

For further details concerning the sensitivity to the delay $\Delta$, the
maximum polynomial order $p$ and the data preparation can be found in the
original papers.\cite{s3,s4}

The reconstructions are seen to be very good and robust for R~Sct.  Col.~4 of
Fig.~6 shows that this remains true when we reconstruct the dynamics with a 4D
flow instead of the 4D map of Col.~2.

The AC~Her data are inferior.  Their signal to noise ratio is much lower and as
a result the reconstruction is much less robust.  It is nevertheless
interesting that we are able to reconstruct a map that can produce synthetic
signals with similar properties to the observational data.

It is not possible to extract Lyapunov exponents and a fractal dimension
directly form the observational data because of their short length.  However we
can obtain these quantities once we have constructed the map or flow.  In Table
1 we show the Lyapunov exponents $\{\lambda_k\}$, ordered from largest to
smallest, and of the Lyapunov dimension $d_L$ for some of the reconstructions
for both R~Sct and AC~Her.  The latter is defined as
 \begin{equation}
 d_L=K+\sum_{i=1}^K\lambda_i/|\lambda_{K+1}| 
 \end{equation}
 where $K$ is the largest value such that the sum is positive.  The
coefficients have been obtained from the maps that have been constructed with
the listed parameters $d_e$, $\Delta$ and $p$.

 \begin{table}
 {\small
 \caption[]
 {
 Lyapunov exponents and Lyapunov dimensions for R~Scuti and AC~Herculis
 }
 \begin{tabular}{ccccccc|ccccccc}
 \noalign{\smallskip}
 \hline\noalign{\smallskip}
 \noalign{\smallskip}
 \noalign{\ \ \ R Scuti \hspace{6.3cm} AC Herculis}
 \noalign{\smallskip}
 \hline\noalign{\smallskip}
 \noalign{\smallskip}
 \ngth${\rm d}_e$&\ngth$\Delta$\ngth&$p$&\ngth$\lambda_1$
 &$\lambda_3$&$\lambda_4$&${\rm d}_{L}$
\ &${\rm d}_{e}\ngth$&\ngth$\Delta$&$p$&$\lambda_1$
 &$\lambda_3$&$\lambda_4$&
${\rm d}_{L}$\cr

 \noalign{\smallskip}
 \hline
 \noalign{\smallskip}
 4&4&\ngth 4& 0.0019 & \ngth --0.0016\ngth&\ngth --0.0061\ngth &\ngth 3.05
  &\ngth 3 &\ngth  5\ngth & 6 & \ngth 0.0033 \ngth&\ngth --0.034\ngth
 &\ngth         & 2.10 \cr
 4&5&\ngth 4& 0.0017 & \ngth --0.0014\ngth&\ngth --0.0054\ngth &\ngth 3.06
  &\ngth 3 &\ngth 10\ngth & 6 & \ngth 0.0045 \ngth&\ngth --0.026\ngth
 &\ngth         & 2.17 \cr
 4&6&\ngth 4& 0.0019 & \ngth --0.0009\ngth&\ngth --0.0051\ngth &\ngth 3.19
  &\ngth 3 &\ngth 13\ngth & 6 & \ngth 0.0069 \ngth&\ngth --0.030\ngth
 &\ngth         & 2.23 \cr
 4&7&\ngth 4& 0.0020 & \ngth --0.0011\ngth&\ngth --0.0052\ngth &\ngth 3.18
  &\ngth 4 &\ngth  5\ngth & 5 & \ngth 0.0073 \ngth&\ngth --0.016\ngth
 &\ngth --0.054 & 2.46 \cr
 4&8&\ngth 4& 0.0014 & \ngth --0.0010\ngth&\ngth --0.0049\ngth &\ngth 3.07
  &\ngth 5 &\ngth  6\ngth & 4 & \ngth 0.0045 \ngth&\ngth --0.023\ngth
 &\ngth --0.025 & 2.19 \cr
 5&7&\ngth 3& 0.0016 & \ngth --0.0005\ngth&\ngth --0.0041\ngth &\ngth 3.27
  &\ngth 5 &\ngth  7\ngth & 4 & \ngth 0.0075 \ngth&\ngth --0.009\ngth
 &\ngth --0.025 & 2.85 \cr
 6&8&\ngth 3& 0.0022 & \ngth --0.0003\ngth&\ngth --0.0018\ngth &\ngth 3.52
  &\ngth 3 &\ngth 13\ngth & 6 & \ngth 0.0015 \ngth&\ngth --0.029\ngth
 &\ngth         & 2.05 \cr 
 \noalign{\smallskip}
 \hline
 \end{tabular}
 }
\end{table}

\section{Results}

We only note the most important points:\hfill\break
 \ni (1) While there is some scatter in the exponents and in the associated
fractal dimension $d_L$, for all successful reconstructions $d_L<4$ for R~Sct
and 3 for AC~Her, independently of the embedding dimension.  This is an
indication that we have achieved an embedding of the dynamics.\hfill\break
 \ni (2) The largest of the Lyapunov exponents is always positive. This is a
confirmation that the signal is indeed chaotic.\hfill\break
 \ni (3) We have not shown the second Lyapunov exponent because it is always
close to zero.  This result is reassuring, because it indicates that we are
actually close to describing a flow for which one of the exponents would be
exactly zero (autonomous system).\hfill\break
 \ni (4) For R~Sct $|\lambda_3|<\lambda_1$, whereas the inequality is reversed
for AC~Her, indicating a larger dissipation in the dynamics of AC~Her.
\hfill\break

As physicists we would like to infer more than just Lyapunov exponents and
fractal dimensions, because as such they are not very useful.  Thus it is
interesting that from the linearization of the R~Sct map about its fixed point
we found\cite{s3} two spiral roots, one unstable and the second stable with
approximately twice the frequency.  This led to the interesting physical
interpretation that the chaotic dynamics is generated by the nonlinear
interaction of two radial modes of pulsation that are in a close resonance
condition.  Such a physical picture is supported by further theoretical
considerations.\cite{s3,mito} As for AC~Her, it is quite likely that two
strongly interacting vibrational modes also underlie this dynamics, but that
the stronger dissipation alluded to just above compresses the dynamics down to
3D, in the same way the 2D H\'enon map shrinks to a 1D logistic map for small
values of $b$.

\section {Conclusions}

The flow reconstruction has been applied to real data, {\sl viz.} the
light-curve of the irregular variable stars R~Scuti and AC~Herculis.  Not only
are these data contaminated with a good amount of observational and other
extrinsic noise, but we have no a priori information about the dynamics that
has generated them.  The analysis has shown that the dynamics which generates
the light-curve of R~Scuti has a dimension of four, i.e. the signal can be
generated by a flow in 4-D.  Furthermore the attractor has a fractal dimension
$\approx $ 3.1.  For the star AC~Her we find a phase-space dimension of 3 and a
fractal dimension $2.05 \approxlt d_L \approxlt 2.5$.

There are no observational data sets available on the even less irregular W~Vir
stars.  However, numerical hydrodynamical simulations and their analysis with
the global flow reconstruction\cite{s2b} indicate also a 3-D phase-space and a
fractal dimension of $d_L\approx$ 2.01 -- 2.05. A systematic trend seems to
exist, {\sl viz.} that as we go from the low luminosity, weakly irregular W~Vir
stars to the high luminosity, strongly irregular RV~Tau stars, the fractal
dimension gradually increases.

The fact that these stars have such a low dimensional dynamics must be
considered remarkable.  They undergo large luminosity fluctuations, up to a
factor of 40 for ~Sct, with strong shock waves and ionization fronts
criss-crossing the stellar envelope.  Yet, the overall pulsational behavior is
found to be relatively simple.

As a result of this nonlinear analysis the nature of the irregular variability
of these types of stars is no longer a mystery, but it has been shown to be
simply the manifestation of an underlying low dimensional chaotic dynamics.

\section*{Acknowledgements}
 It is a great pleasure to acknowledge the contributions of my collaborators,
Zoltan Koll\'ath and Thierry Serre.  This work has been supported by
NSF~(AST95--18068).


\begin{thebibliography}{99}

{

 \bibitem{PDletter} 
 J. R. Buchler \& G. Kov\'acs, {\sl Astrophys. J. Lett.} {\bfrm 320} (1987)
L57.

 \bibitem{PDpaper} G. Kov\'acs \& J. R. Buchler, {\sl Astrophys. J.}  {\bfrm
334} (1988) 971.

 \bibitem{PRlett} 
  J. R. Buchler, T. Serre, Z. Koll\'ath \& J. Mattei, {\sl Phys. Rev.  Lett.}
{\bfrm 74} (1995) 842.

 \bibitem{s3}
 J. R. Buchler, Z. Koll\'ath, T. Serre, \& J. Mattei., {\sl Astrophys. J.},
{\bfrm 462} (1996) 489 astro-ph/9707116.

 \bibitem{s4} 
 Z. Koll\'ath, J. R. Buchler, \& T. Serre, {\sl Astr. \& Astrophys.}  (1997)
(in press) astro-ph/9707099.

\bibitem{Abar}
  Abarbanel, H.D.I., Brown, R.,  Sidorowich, J. J.,  Tsimring, L. S.
{\sl Rev. Mod. Phys.} {\bfrm 65} (1993)  1331.

 \bibitem{WG}
 Weigend, A.S \& Gershenfeld, N. A.  1994, {\sl Time Series Prediction}
(Addison-Wesley: Reading).

 \bibitem{Sauer}
  Sauer, T., Yorke, J. A. \& Casdagli, M., {\sl J. Stat. Phys.}  (1991) {\bfrm
65}, 5.

 \bibitem{Giona}
 Giona, M., Lentini, F. \& Cimagalli, V., {\sl Phys. Rev.} {\bfrm A 44} (1991) 
3496.

 \bibitem{Brown}
 Brown, R. 1992, {\sl Orthonormal Polynomials As Prediction Functions In
Arbitrary Phase-Space Dimensions}, Institute for Nonlinear Science Preprint, UC
San Diego.

 \bibitem{s2a} 
 T. Serre, Z. Koll\'ath, \& J. R. Buchler, {\sl Astr. \& Astrophys.}
{\bfrm 311} (1996) 833.

 \bibitem{roessler}
 R\"ossler, O. E., {\sl Phys. Lett.} {\bfrm 57A} (1976) 397.

 \bibitem{s2b}
 T. Serre, Z. Koll\'ath, \& J. R. Buchler, {\sl Astr. \& Astrophys.} {\bfrm
 311} (1996) 845.

 \bibitem{mito}
 Buchler, J. R. 1993, in {\sl Nonlinear Phenomena in Stellar Variability},
Eds. M. Takeuti \& J.R. Buchler (Kluwer: Dordrecht), repr. from {\sl
Ap\&SS} 210 (1993) 1.

 } 
 \end{thebibliography}
 \end{document}